\documentclass[aps,prb,preprint,showpacs,showkeys,floatfix]{revtex4}

\usepackage{graphicx,color}
\usepackage{multirow,slashbox}

\graphicspath{{figs/}}
\begin{document}
\title{Machine learning valence force field model}

\author{Jing Wan}
    \affiliation{Shanghai Institute of Applied Mathematics and Mechanics, Shanghai Key Laboratory of Mechanics in Energy Engineering, Shanghai University, Shanghai 200072, People's Republic of China}

\author{Ya-Wen Tan}
    \affiliation{Shanghai Institute of Applied Mathematics and Mechanics, Shanghai Key Laboratory of Mechanics in Energy Engineering, Shanghai University, Shanghai 200072, People's Republic of China}

\author{Jin-Wu Jiang}
    \altaffiliation{Corresponding author: jiangjinwu@shu.edu.cn; jwjiang5918@hotmail.com}
    \affiliation{Shanghai Institute of Applied Mathematics and Mechanics, Shanghai Key Laboratory of Mechanics in Energy Engineering, Shanghai University, Shanghai 200072, People's Republic of China}

\author{Tienchong Chang}
    \affiliation{Shanghai Institute of Applied Mathematics and Mechanics, Shanghai Key Laboratory of Mechanics in Energy Engineering, Shanghai University, Shanghai 200072, People's Republic of China}

\author{Xingming Guo}
\affiliation{Shanghai Institute of Applied Mathematics and Mechanics, Shanghai Key Laboratory of Mechanics in Energy Engineering, Shanghai University, Shanghai 200072, People's Republic of China}

\date{\today}
\begin{abstract}

The valence force field (VFF) model is a concise physical interpretation of the atomic interaction in terms of the bond and angle variations in the explicit quadratic functional form, while the machine learning (ML) method is a flexible numerical approach to make predictions based on some pre-obtained training data without the need of any explicit functions. We propose a so-called ML-VFF model, by combining the clear physical essence of the VFF model and the numerical flexibility of the ML method. Instead of imposing any explicit functional forms for the atomic interaction, the ML-VFF model predicts the potential and force with the Gaussian regression approach. We take graphene as an example to illustrate the ability of the ML-VFF model to make accurate predictions with relatively low computational expenses. We also discuss some key advantages and drawbacks of the ML-VFF model.

\end{abstract}

\keywords{Machine learning, potential, graphene}
\maketitle
\pagebreak

\section{Introduction}

The atomic interaction plays a fundamental role in governing both static and dynamic progresses for materials on the microscopic atomic level. The first-principles calculations can provide accurate atomic interactions for most materials. However, the first-principles calculations demand rather expensive computational costs for large systems, where empirical potentials are desirable. As a consequence, a large number of empirical potentials have been proposed, including the valence force field (VFF) model, the Stillinger-Weber potential,\cite{StillingerF} the Tersoff potential,\cite{TersoffJ1} and the Brenner potential,\cite{brennerJPCM2002} etc. These empirical potentials have explicit functional forms that are constructed in the spirit of some physical laws. The accuracy of the empirical potential is closely related to the complexity of its functional form. A more complex function can contain more physical effects, so is usually more accurate.

Basically, the task of the empirical potential is to provide (predict) the atomic potential and the atomic force for an arbitrary given structure. The empirical potential is very efficient in such predicting, owing to its explicit functional form. For instance, the VFF model is a linear model with the simplest quadratic function of the bond length $b$ and bond angle $\theta$, $V=V(b,\theta)=\frac{1}{2}K_b(\Delta b)^2+\frac{1}{2}K_{\theta}(\Delta \theta)^2$, where $K_b$ and $K_{\theta}$ are the corresponding force constants. With this quadratic functional form, the VFF model treats bonds and angles as two kinds of linear springs. The potential and force of a given structure can thus be predicted analytically by the VFF model.

In terms of prediction, there are some machine learning (ML) approaches, like the Gaussian regression method, that are good at making predictions based on some available data. Taking advantage of its predictability, the ML approach has been applied to investigate some physical or mechanical processes, such as the modeling of molecular atomization energies,\cite{RuppM2012prl} the energies of elpasolite (ABC$_2$D$_6$) crystals,\cite{FaberFA2016prl} topological invariants,\cite{ZhangP2018prl} defect and phase evolution during transformations in WS$_2$,\cite{MaksovA2018arxiv} vibrational properties of metastable polymorph structures,\cite{LegrainF2018arxiv} and thermoelectric materials.\cite{IwasakiY2018arxiv} In particular, the ML approach can be utilized to predict the atomic potential and atomic force using the artificial neural networks\cite{BehlerJ2007prl,KhaliullinRZ2010prb,GhasemiSA2015prb} or the Gaussian approximation,\cite{BartokAP2010prl,BartokAP2015ijqc,BartokAP2017sa,ArtrithN2017prb} which have been applied to tantalum,\cite{ThompsonAP2015jcp} amorphous carbon,\cite{DeringerVL2017prb} graphene,\cite{RoweP2018prb} amorphous silicon,\cite{DeringerVL2018arxiv} silicon,\cite{BartokAP2018arxiv} boron,\cite{DeringerVL2018prl} and boron carbide,\cite{GaoQ2015arxiv} etc.

From the above, an explicit functional form is a characteristic feature that enables the high efficiency and clear physics for the empirical potential. However, the explicit functional form also causes some limitations for the empirical potential, because the functional form is constructed based on few physical considerations and will inevitably lose some other physical information. For example, the VFF model treats the bond and angle as linear springs, so nonlinear effects are ignored. On the other hand, the ML approach is a purely numerical or mathematical approach, which does not include any physical effects; i.e., no explicit functional form is required.\cite{KurkovaV1992nn} As a consequence of its numerical nature, the ML approach is highly flexible in transferring between different materials, because the ML approach is not constrained by any physical laws in different materials.

We are now aware that the VFF model contains concise physical essences, while the ML approach is extremely flexible. We try to combine these two advanced features. A minimum physical consideration is to assume that the atomic interaction is a function of the bond length and bond angle, $V=V(b,\theta)$. In contrast to the usual VFF model, the functional form for $V(b,\theta)$ is not assumed, but the atomic potential and force are predicted by the ML approach. In doing so, we actually generalize the linear VFF model to be a nonlinear model, including all high-order nonlinear terms, so this new model will be referred to as the ML-VFF model.

In this paper, by combining the advantages of the VFF model and the ML technique, we propose the ML-VFF model to describe the atomic interactions. In contrast to the usual VFF model that is a quadratic function of the bond and angle variations, the ML-VFF model does not have an explicit functional form, while the potential and force are predicted by the Gaussian regression method. The ML-VFF model is able to approach the accuracy of other more complex models with much lower computational expenses. We illustrate the training of the ML-VFF model to the Brenner potential in graphene, and demonstrate the accuracy and efficiency of the ML-VFF model in the simulations.

\section{VFF model and Brenner potential}

In relevance to the present work, the VFF model and the Brenner potential are two typical empirical potential models that have been widely used in the simulation of solid materials. They represent two extreme situations in developing empirical models for the atomic potential. Bonds and angles are represented by linear springs in the VFF model. The VFF model is the simplest model, so it is the most efficient but less accurate. The Brenner potential has a much more complex functional form, so it is more accurate but less efficient.

The VFF model assumes that the atomic interaction is a function of the bond variation $\Delta b$ and the angle variation $\Delta\theta$. It is further assumed that the atomic interaction takes the following quadratic functional form,
\begin{eqnarray}
V_{b} & = & \frac{1}{2}K_{b}\left(\Delta b\right)^{2},\\
\label{eq_vffm1}
V_{\theta} & = & \frac{1}{2}K_{\theta}\left(\Delta\theta\right)^{2},
\label{eq_vffm2}
\end{eqnarray}
where $K_{b}$ and $K_{\theta}$ are two parameters. By taking this quadratic functional form, the VFF model is constrained in the linear regime, so it is of high efficiency in computation, but this model is typically less accurate as a trade off. The underlying mechanism for the VFF model is that materials are discretized by bonds and angles that are treated as linear springs.

The Brenner potential is more complex than the VFF model. The atomic interaction is assumed to be a function of the atom's bond-order environment in the Brenner model, and the explicit function has a much more complex form. Specifically, the energy in the Brenner potential is 
\begin{eqnarray}
V_{B}(r) & = & V_{R}(r)-\bar{B}_{ij}\cdot V_{A}(r).
\end{eqnarray}
$V_{R}$ and $V_{A}$ are the repulsive and attractive energy,
\begin{eqnarray}
V_{R}(r) & = & \frac{D^{(e)}}{S-1}e^{-\sqrt{2S}\beta(r-R^{(e)})}f_{c}(r)\\
V_{A}(r) & = & \frac{D^{(e)}S}{S-1}e^{-\sqrt{2/S}\beta(r-R^{(e)})}f_{c}(r),
\end{eqnarray}
with the cut-off function $f_{c}(r)$,
\begin{eqnarray}
f_{c}(r) & = & \left\{
\begin{array}{cc}
1, &  \hspace{0.4cm}  r<R^{(1)}\; ,\\
\frac{1}{2}\left\{ 1+\cos\left[\frac{\pi(r-R^{(1)})}{R^{(2)}-R^{(1)}}\right]\right\}, &  \hspace{0.4cm} R^{(1)}<r<R^{(2)}\; ,\\
0, &  \hspace{0.4cm}  r>R^{(2)}\; ,\\
\end{array}
\right.
\end{eqnarray}
where $r$ is the distance between two atoms. The many-body coupling parameter is
\begin{eqnarray}
\bar{B}_{ij} & = & \frac{1}{2}(B_{ij}+B_{ji})\\
B_{ij} & = & \left[1+\sum_{k\not=ij}G(\theta_{ijk})f_{c}(r_{ik})\right]^{-\delta}.
\end{eqnarray}
The angle function $G(\theta_{ijk})$ is
\begin{eqnarray}
G(\theta_{ijk}) & = & a_{0}\left[1+\frac{c_{0}^{2}}{d_{0}^{2}}-\frac{c_{0}^{2}}{d_{0}^{2}+\left(1+\cos(\theta_{ijk})\right)^{2}}\right]
\end{eqnarray}
where $\theta_{ijk}$ is the angle formed by atoms i, j, and k. All parameters in the Brenner potential can be found in Ref.~\onlinecite{brennerJPCM2002} for carbon materials.

Both the VFF model and the Brenner potential have explicit functional forms, which are constructed based on particular physical essences. The atomic potential and the atomic force can be predicted in an efficient manner by these explicit functions. However, in the meantime, these explicit functional forms will inevitably miss some other physics, which may be important in some materials.

\section{ML-VFF model}

Similar as the usual VFF model, the atomic interaction is also assumed to be a function of the bond and the angle in the ML-VFF model. However, different from the usual VFF model, no explicit functional form is assumed for the ML-VFF model, because the ML approach can make predictions without explicit functional forms.\cite{KurkovaV1992nn}

Similar as the GAP model,\cite{RoweP2018prb} the total potential is decomposed into atomic potential, which is a collection of the contribution from different Kernel functions. Specifically, the potential energy for atom i in the ML-VFF model is,
\begin{eqnarray}
V_{i}^{ml} & = & \sum_{p=1}^{N_{b}}\alpha_{p}^{b}\sum_{<ij>}K_{pi}^{b}+\sum_{p=1}^{N_{\theta}}\alpha_{p}^{\theta}\sum_{<ijk>}K_{pi}^{\theta}
\label{eq_Vi}
\end{eqnarray}
where the superscript ml indicates ``machine learning". The summation $<ij>$ runs over all neighbors for atom i, and $<ijk>$ runs over all angles for atom i. $N_b$ and $N_{\theta}$ are the number of sparse points for the bond and angle spaces, respectively. $\alpha^b$ and $\alpha^{\theta}$ are the coefficients that will be determined in the training process. $K_{pi}^{b}$ and $K_{pi}^{\theta}$ are the kernel functions, which are chosen to be the following Gaussian kernel (also called the squared exponential kernel) in the present work,
\begin{eqnarray}
K_{pi}^{b} & = & e^{-\frac{\left(r_{ij}-b_{p}\right)^{2}}{2\xi_{b}^{2}}}\\
K_{pi}^{\theta} & = & e^{-\frac{\left(\theta_{ijk}-\theta_{p}\right)^{2}}{2\xi_{\theta}^{2}}}
\end{eqnarray}
where $\xi_b$ and $\xi_{\theta}$ are two parameters.

As a result, the total energy for a given structure is
\begin{eqnarray}
V^{ml} & = & \sum_{i=1}^{N}V_{i}^{ml}\nonumber\\
 & = & \sum_{p=1}^{N_{b}}\alpha_{p}^{b}\left(\sum_{i=1}^{N}\sum_{<ij>}K_{pi}^{b}\right)+\sum_{p=1}^{N_{\theta}}\alpha_{p}^{\theta}\left(\sum_{i=1}^{N}\sum_{<ijk>}K_{pi}^{\theta}\right)
\label{eq_Vtot}
\end{eqnarray}
where $N$ is the number of atoms in the structure. The atomic force for atom i can be derived from the atomic potential in Eq.~(\ref{eq_Vi}),
\begin{eqnarray}
\vec{F}_{i}^{ml} & = & -\nabla V_{i}^{ml}\nonumber\\
 & = & -\sum_{p=1}^{N_{b}}\alpha_{p}^{b}\left(\sum_{<ij>}\nabla K_{pi}^{b}\right)-\sum_{p=1}^{N_{\theta}}\alpha_{p}^{\theta}\left(\sum_{<ijk>}\nabla K_{pi}^{\theta}\right).
\label{eq_fi}
\end{eqnarray}

To determine the coefficients $\alpha^b$ and $\alpha^{\theta}$ in the training process, we point out that, the total potential and the atomic force are linear functions of the coefficients $\alpha^{b}$ and $\alpha^{\theta}$ as shown in Eqs.~(\ref{eq_Vtot}) and (\ref{eq_fi}). Hence, it is straightforward to determine these coefficients $\alpha^{b}$ and $\alpha^{\theta}$ by the linear least-squares fit. The merit function for the linear least-squares fit is defined as
\begin{eqnarray}
\chi^{2} & = & \sum_{t=1}^{N_{t}}\left(\frac{V_{t}-V_t^{ml}}{\sigma_{t}}\right)^{2}+\sum_{t=1}^{N_{t}}\sum_{i=1}^{N}\left(\frac{\vec{F}_{ti}-\vec{F}_{ti}^{ml}}{\vec{\delta}_{ti}}\right)^{2}
\label{eq_chi2}
\end{eqnarray}
where $N_t$ is the total number of structures in the training set. $V_t$ is the total potential energy for the training structure t with $\sigma_t$ as the error. $\vec{F}_{ti}$ is the atomic force for atom i in the training structure t and $\vec{\delta}_{ti}$ is the error. These quantities ($V_t$, $\sigma_t$) and ($\vec{F}_{ti}$, $\vec{\delta}_{ti}$) are input as training data, which have been pre-obtained experimentally or theoretically. These two quantities $V_t^{ml}$ and $\vec{F}_{ti}^{ml}$ are calculated from Eqs.~(\ref{eq_Vtot}) and (\ref{eq_fi}) for the training structure t. In the training process, the coefficients $\alpha^{b}$ and $\alpha^{\theta}$ are determined by minimizing the merit function in Eq.~(\ref{eq_chi2}).

In the present ML-VFF model, these two physical quantities, bond length and bond angle, are used as the variables for the unknown function of the atomic potential, leading to a natural generalization of the standard VFF model. The bond length and bond angle have exact physical meaning, but they are not complete for the description of the whole crystal structure due to their short-ranged nature. For a fully description of the crystal structure, readers are referred to some other more complete quantities, such as the smooth overlap of atomic positions,\cite{BartokAP2010prl} atom-centered symmetry functions,\cite{BehlerJ2011jcp} and others discussed in the review article.\cite{BartokAP2013prb}

\section{Application to graphene}

We will now take graphene as an explicit example to illustrate how to train and use the ML-VFF model.

\subsection{Training ML-VFF model}

\begin{figure}[tb]
  \begin{center}
    \scalebox{0.7}[0.7]{\includegraphics[width=8cm]{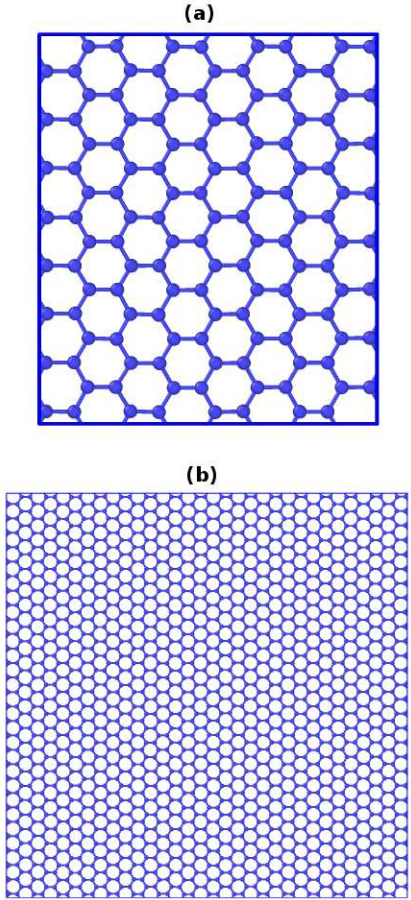}}
  \end{center}
  \caption{(Color online) Graphene sheets with (a) 128 carbon atoms and (b) 1792 carbon atoms. The smaller structure is used for training, while the prediction is made for the larger structure.}
  \label{fig_cfg}
\end{figure}

\begin{table}
\caption{Parameters used for training the ML-VFF model of graphene. The length is in the unit of $\AA$, while the angle is in the unit of rad.}
\label{tab_parameter}
\begin{tabular}{@{\extracolsep{\fill}}|c|c|c|}
\hline 
parameter & meaning & value\tabularnewline
\hline 
\hline 
($b_{\rm min}$, $b_{\rm max}$) & boundaries for bond space & (1.0, 4.0)\tabularnewline
\hline 
($\theta_{\rm min}$, $\theta_{\rm max}$) & boundaries for angle space & (1.75, 2.45)\tabularnewline
\hline 
$N_{b}$ & grid number for bond space & 10\tabularnewline
\hline 
$N_{\theta}$ & grid number for angle space & 10\tabularnewline
\hline 
$N_{t}$ & number of training set & 2000\tabularnewline
\hline 
$N$ & number of atom in a training structure & 128\tabularnewline
\hline 
$\xi_{b}$ & decaying factor in the kernel function for bond & 1.0\tabularnewline
\hline 
$\xi_{\theta}$ & decaying factor in the kernel function for angle & 1.0\tabularnewline
\hline 
\end{tabular}
\end{table}

\begin{table}
\caption{Training results for the ML-VFF model of graphene. The length is in the unit of $\AA$, while the angle is in the unit of rad.}
\label{tab_alpha}
\begin{tabular}{@{\extracolsep{\fill}}|c|c|c|c|c|}
\cline{1-2} \cline{4-5} 
bond $b_{p}$  & coefficient $\alpha_{p}^{b}$ & \multirow{11}{*}{} & angle $\theta_{p}$ & coefficient $\alpha_{p}^{\theta}$\tabularnewline
\cline{1-2} \cline{4-5} 
1.00 & 8131.39 &  & 1.75 & -494.54\tabularnewline
\cline{1-2} \cline{4-5} 
1.33 & -3079.30 &  & 1.83 & -16542.84\tabularnewline
\cline{1-2} \cline{4-5} 
1.67 & 2571.36 &  & 1.91 & 11746.63\tabularnewline
\cline{1-2} \cline{4-5} 
2.00 & -23794.57 &  & 1.98 & -7043.63\tabularnewline
\cline{1-2} \cline{4-5} 
2.33 & 16033.22 &  & 2.06 & 6610.09\tabularnewline
\cline{1-2} \cline{4-5} 
2.67 & 35090.30 &  & 2.14 & 17345.95\tabularnewline
\cline{1-2} \cline{4-5} 
3.00 & -16715.71 &  & 2.22 & -5845.43\tabularnewline
\cline{1-2} \cline{4-5} 
3.33 & -33813.44 &  & 2.29 & 2715.09\tabularnewline
\cline{1-2} \cline{4-5} 
3.67 & -27394.05 &  & 2.37 & -5951.99\tabularnewline
\cline{1-2} \cline{4-5} 
4.00 & 76057.14 &  & 2.45 & -5393.47\tabularnewline
\cline{1-2} \cline{4-5} 
\end{tabular}
\end{table}

The ML-VFF model is trained to the Brenner potential in graphene. Molecular dynamics (MD) simulations are performed to generate the training data. The interaction in graphene is described by the Brenner potential.\cite{brennerJPCM2002} The standard Newton equations of motion are integrated in time using the velocity Verlet algorithm with a time step of 1~{fs}. A small piece of graphene with 128 carbon atoms as shown in Fig.~\ref{fig_cfg}~(a) is biaxially stretched at 100~K temperature. The structure is stretched by the strain of 0.05 after 10000 simulation steps. From the MD simulation, we thus obtain 10000 instant structures, from which we have randomly selected 2000 structures as the training set while another 1000 structures as the testing set. There are in total 385 data in each training structure, including 1 total potential and 384 atomic forces for the graphene sheet.

The ML-VFF model is trained to the training set containing 2000 structures with parameters listed in Tab.~\ref{tab_parameter}. The four boundaries $b_{\rm min}$, $b_{\rm max}$, $\theta_{\rm min}$, and $\theta_{\rm max}$ are chosen to include the value of the equilibrium bond lengths (around 1.42~{\AA}) and bond angles (around $120^{\circ}$) of graphene. The two decaying factors $\xi_b$ and $\xi_{\theta}$ are chosen to be on the same order as $(b_{\rm max}-b_{\rm min})$ and $(\theta_{\rm max}-\theta_{\rm min})$, respectively; while some variations in these two decaying factors do not affect the training results. The results are not sensitive to these numbers $N_b$, $N_{\theta}$, $N_t$, and $N$.

\begin{figure}[htpb]
  \begin{center}
    \scalebox{1}[1]{\includegraphics[width=8cm]{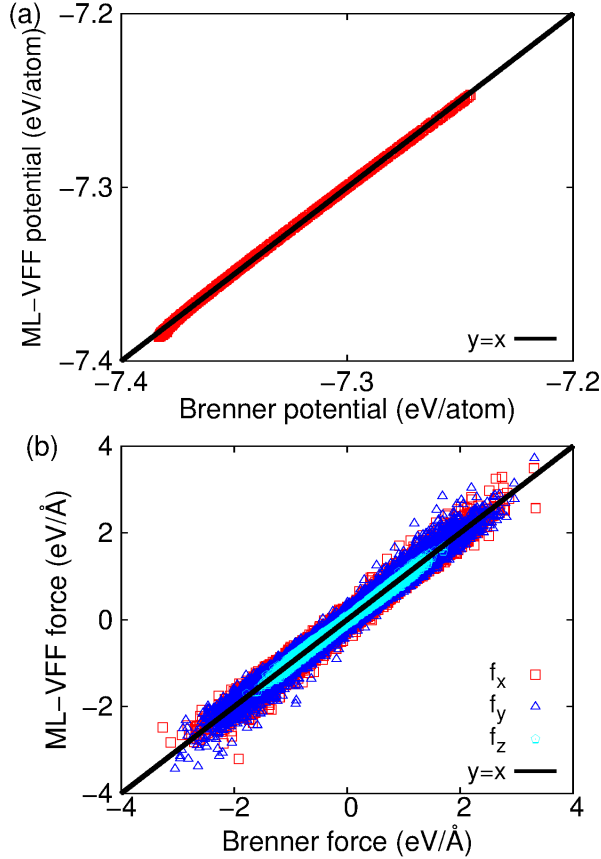}}
  \end{center}
  \caption{(Color online) Test the accuracy of the ML-VFF model with one thousand structures. (a) The total potential energy predicted by the ML-VFF model versus that calculated from the Brenner potential for these testing structures. (b) The atomic force predicted by the ML-VFF model versus that calculated from the Brenner potential for these testing structures.}
  \label{fig_test}
\end{figure}

The key objective of the training process is to determine the coefficients $\alpha_b$ and $\alpha_\theta$, as listed in Tab.~\ref{tab_alpha}. With the trained coefficients $\alpha_b$ and $\alpha_\theta$, the ML-VFF model can then be applied to calculate the total potential and the atomic force for the 1000 structures in the testing set according to Eqs.~(\ref{eq_Vtot}) and (\ref{eq_fi}). The total potential and the atomic force for the testing set can also be computed by using the Brenner potential. Results from the Brenner potential and the ML-VFF model are compared in Fig.~\ref{fig_test}.

Figure~\ref{fig_test}~(a) displays the relation between the potential for the testing structures calculated from the Brenner potential and that from the ML-VFF model. There are 1000 data in the figure. A close matching between these numerical data (red online) and the solid line (black online) of $y=x$ indicates that the ML-VFF model reproduces quite well the Brenner potential of the 1000 testing graphene structures.

Figure~\ref{fig_test}~(b) illustrates the relationship between the atomic force calculated from the Brenner potential and that from the ML-VFF model. There are totally 384000 data plotted in the figure. It shows that the force for each carbon atom in the 1000 testing structures is also well reproduced by the ML-VFF model. In particular, the usual VFF model is not able to generate force in the out-of-plane direction ($f_z$) of graphene, because of its quadratic functional expression;\cite{JiangJW2008} while the ML-VFF model can produce $f_z$ of high accuracy.

\subsection{Prediction from ML-VFF model}

\begin{figure}[htpb]
  \begin{center}
    \scalebox{1}[1]{\includegraphics[width=8cm]{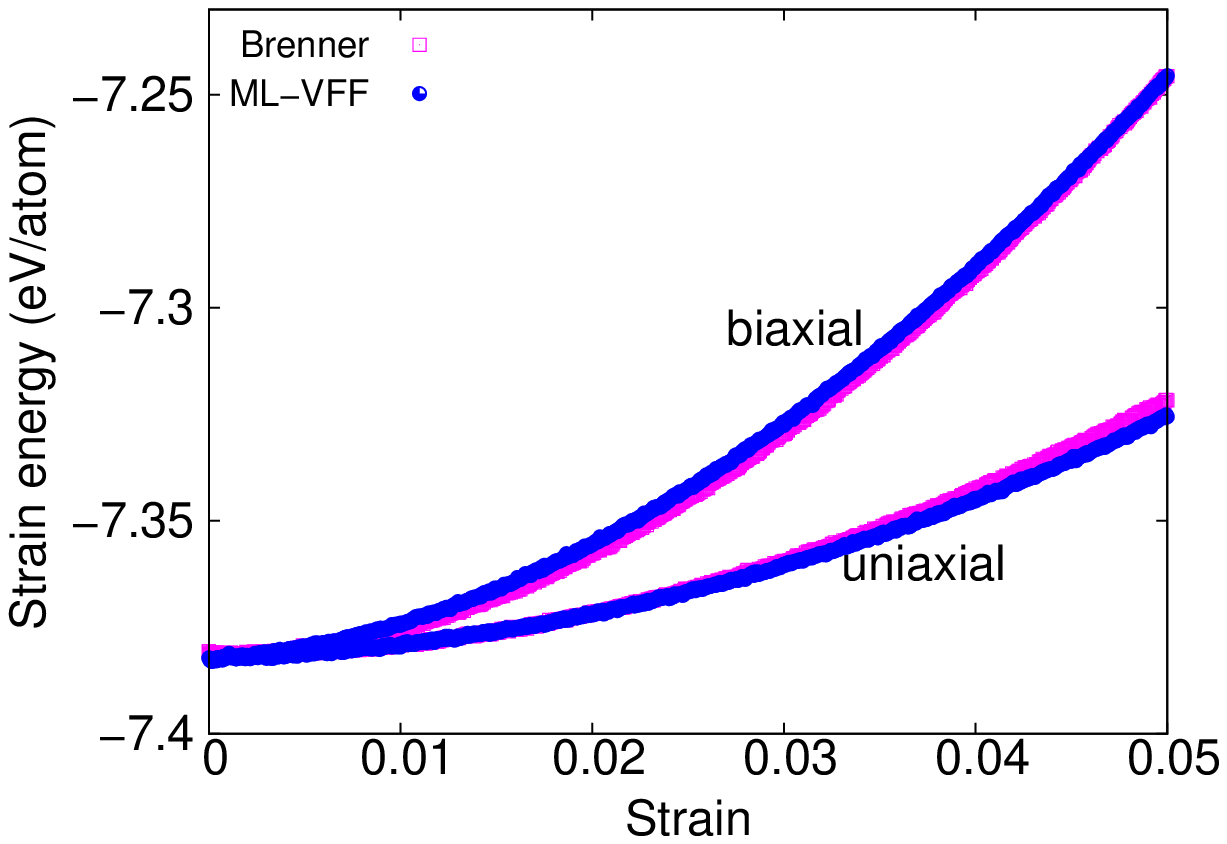}}
  \end{center}
  \caption{(Color online) The strain energy versus the strain for the biaxial and uniaxial stretching of graphene with 1792 carbon atoms.}
  \label{fig_strain_energy}
\end{figure}

\begin{figure}[htpb]
  \begin{center}
    \scalebox{1}[1]{\includegraphics[width=8cm]{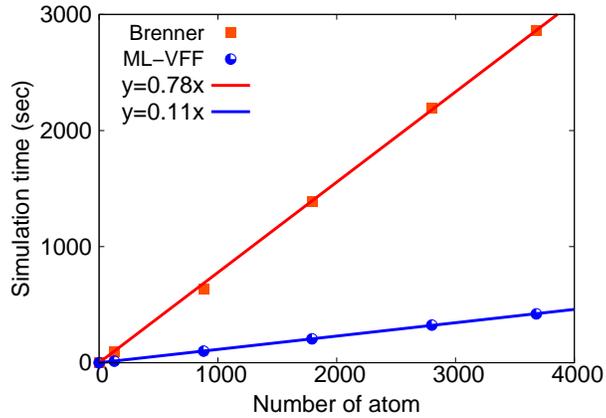}}
  \end{center}
  \caption{(Color online) Comparison of the simulation time between Brenner potential and the ML-VFF model.}
  \label{fig_time}
\end{figure}

The ML-VFF model can be applied to simulate the stretching for graphene sheet of arbitrary size, though it is trained with small pieces of graphene structures. We stretch a larger graphene sheet with 1792 carbon atoms at 100~K temperature as shown in Fig.~\ref{fig_cfg}~(b). Fig.~\ref{fig_strain_energy} shows the strain energy for biaxial and uniaxial stretching of the structure that is computed based on either the Brenner potential or the ML-VFF model. We find that the ML-VFF model gives the same strain energies as the Brenner potential for the biaxial and uniaxial stretching processes. In other words, the ML-VFF model is able to make reliable predictions for the stretching process in graphene sheets of larger size.

The power of the ML-VFF model is to save lots of computational cost, while keeping the accuracy not affected. To illustrate this ability, we compare the simulation cost for the Brenner potential and the ML-VFF model in Fig.~\ref{fig_time}. MD simulations are performed for five structures with the number of atoms as 128, 880, 1792, 2800, and 3680, where the atomic interaction is described by either the Brenner potential or the ML-VFF model. All MD simulations are running for 10000 steps. The simulation time increases linearly with the increase of the number of atom in the structure. The simulation time per atom is 0.78~second for the Brenner potential, and 0.11~second for the ML-VFF model. Hence, the simulation cost for the ML-VFF model is only about one seventh of the Brenner potential, while they have the same accuracy.

Furthermore, the ML-VFF model is quite flexible. There is no assumption for the functional form of the atomic interaction, so this model is not restricted by any physical effects and is applicable to all kinds of materials. One only needs to provide the training data for the corresponding materials that they are interested in.

\subsection{Limitation of ML-VFF model}

We have demonstrated the accuracy and efficiency for the ML-VFF model in the above. However, there is one common drawback in all ML based predictions, including the ML-VFF model developed in the present work. \textit{The common drawback is that the accuracy of the ML based prediction is dependent on the completeness of the information contained in the training set.} For instance, no defect has been included in the above training data, so there is no guarantee that the ML-VFF model trained to this training set can make sound predictions for the creation of defects. To resolve this issue, one needs to add some configurations with defects into the training set.

\section{Conclusion}

To summarize, we have proposed a numerical version of the VFF model based on the ML approach, which is referred to as the ML-VFF model. In contrast to the usual VFF model, the ML-VFF model does not have an explicit functional form, while the potential and force for a given structure are predicted by the Gaussian regression approach. The ML-VFF is applicable to describe the interaction for materials, where the interaction can be regarded as a function of the bond length and angle. As an example, we train the ML-VFF model to the Brenner potential of graphene, and apply the resultant ML-VFF model to simulate the mechanical loading process of graphene. It is shown that the results from the ML-VFF model can be as accurate as the Brenner potential at much lower computation expense. We discuss some key advantages and drawbacks for the ML-VFF model in practical applications.

\textbf{Acknowledgements} The authors thank G\'abor Cs\'anyi at the University of Cambridge for valuable communications. The work is supported by the Recruitment Program of Global Youth Experts of China, the National Natural Science Foundation of China (NSFC) under Grant No. 11504225, and the Innovation Program of Shanghai Municipal Education Commission under Grant No. 2017-01-07-00-09-E00019.

\textbf{Competing financial interests} The authors declare no competing financial interests.


\end{document}